\documentclass[preprintnumbers,superscriptaddress,amsmath,amssymb,prd,nofootinbib,onecolumn]{revtex4-2}
\usepackage{amssymb}
\usepackage{mathrsfs}
\usepackage{amsthm}
\usepackage{braket}
\usepackage{bm}
\usepackage{qcircuit}
\usepackage{graphicx}
\usepackage{amsfonts}
\usepackage[colorlinks,linkcolor=red,anchorcolor=blue,citecolor=green]{hyperref}
\usepackage{subfigure}
\usepackage{float}
\usepackage{enumerate}
\usepackage{multirow}

\usepackage{array}
\usepackage[figuresright]{rotating}
\usepackage{upgreek}

\begin{document}
	
\title{ Higher order analysis of the gravitational wave velocity memory effect between two free-falling gyroscopes  in the plane wave spacetime}

		\author{Yingxin Chen}
	\affiliation{Department of Physics, Shanghai Normal University,
		100 Guilin Rd, Shanghai 200234, P.R.China}
	
		\author{Ke Wang}
	\affiliation{Department of Physics, Shanghai Normal University,
		100 Guilin Rd, Shanghai 200234, P.R.China}

	\author{Chao-Jun Feng}
	\thanks{Corresponding author}
	\email{fengcj@shnu.edu.cn}
	\affiliation{Department of Physics, Shanghai Normal University,
		100 Guilin Rd, Shanghai 200234, P.R.China}

	\begin{abstract}
		In the plane wave spacetime, when gravitational waves pass by, an angular deviation exists between two free-falling gyroscopes, which naturally corresponds to the velocity memory effect.
		In the shackwave spacetime background, the angular deviation between two free-falling gyroscopes is calculated, which is also found to correspond to the velocity memory effect.
		In the plane wave spacetime, with linear polarization taken into account, no contribution is made by the first-order terms of the initial separation distance \(L\) (or the initial separation velocity \(v_0\)),
		\(\bm{P}\) (or \(\bm{M}\)) of two free-falling gyroscopes to the velocity memory effect, while contributions are initiated from the second-order terms.
		Under certain circumstances, the second-order contribution of the initial separation distance \(L\) is of the same order of magnitude as the first-order contribution of the initial separation velocity \(v_0\).
		When both + polarization and \(\times\) polarization are taken into account, in the context of the merger of supermassive black holes and with the initial separation velocity approaching the speed of light, the order of magnitude of the angle is \(10^{-16}\) rads.
	\end{abstract}

	\maketitle

	\section{Introduction}
	Gravitational wave memory arises from the non-oscillating components of gravitational waves \cite{wang2023spin}, with rotating objects playing a crucial role in detecting the gravitational wave memory effect.
	The known memory effects include the displacement memory effect \cite{zel1974radiation, christodoulou1991nonlinear, blanchet1992hereditary}, the velocity memory effect \cite{grishchuk1989gravitational, zhang2018velocity}, and the gyroscope memory effect \cite{herrera2000influence, seraj2023gyroscopic}, among others.
	When a gravitational wave passes through, the displacement or velocity of freely falling test particles is permanently altered, a phenomenon known as the displacement memory effect or the velocity memory effect.
	If spin is imparted to the freely falling test particles, the gravitational wave induces precession in the rotating particles, permanently changing their orientation. This phenomenon is referred to as the gyroscope memory effect \cite{faye2024gyroscopic}.In \cite{seraj2023gyroscopic}, the motion of a gyroscope in asymptotically flat spacetime far from an isolated gravitational source is studied, generalizing the Lense-Thirring precession to radiative metrics. In \cite{wang2023spin}, the authors examine the precession of a gyroscope in plane wave spacetime. The results indicate that the rotating carrier component associated with the local tetrad remains unchanged, and after the gravitational wave passes through, an angular deviation develops between two freely falling gyroscopes. It is noteworthy that the angular deviation between the two gyroscopes corresponds naturally to the velocity memory effect in plane gravitational wave spacetime.
	
	Given that the back-reaction of detectors to gravitational waves is negligible \cite{zhang2017soft}, a more straightforward method was introduced in \cite{duval2017carroll}, which approximates gravitational waves near the detectors as precise plane waves at a distance from the source. Since gravitational waves reaching Earth are extremely faint, we typically employ the weak-field approximation for their analysis. However, the coefficients of higher-order corrections can grow significantly with the distance between the source and the observer, making nonlinear effects eventually become important. Furthermore, nonlinear effects often produce observational signatures that differ drastically from those of linear effects \cite{harte2015optics}. Therefore, it is essential to explore the exact solutions of Einstein’s equations—nonlinear plane waves \cite{audagnotto2024dynamics}.
	
	The linear memory effect arises from changes in mass or momentum due to a gravitational wave source, while the memory effect caused by the energy flow of the gravitational waves themselves is referred to as the nonlinear memory effect.
	Since the energy of gravitational radiation is second-order in gravitational perturbations, it is clear that nonlinear memory cannot be treated within first-order perturbation theory, leaving second-order perturbation theory as the appropriate framework for its treatment \cite{christodoulou1991nonlinear, bieri2014perturbative}.
	To simplify the problem, one can generally focus on linear polarization, with the other polarization mode made zero through a coordinate transformation. This approach aids in understanding the propagation and effects of gravitational waves in highly symmetric systems, such as spinless binary star systems.
	However, not all scenarios allow the elimination of the other polarization mode via coordinate transformations. For example, in spinning binary star systems, there is also a memory effect for the x-polarization \cite{talbot2018gravitational}.

This study is framed within the context of exact plane waves. In \cite{wang2023spin}, the velocity memory effect is investigated by examining the angular deviation between two gyroscopes. Using a simplified gravitational collapse model with a single polarization, it is shown that no first-order angular deviation occurs. Numerical simulations indicate that, for the sandwich wave, the memory effect only appears in the higher-order terms. 
It is well understood that, under standard assumptions, linear gravitational waves do not produce a final velocity memory effect, though displacement memory is present. The manifestation of the velocity memory effect requires the inclusion of higher-order nonlinear contributions, as confirmed by \cite{wang2023spin}.
In this paper, we derive the general expressions for \(\bm{P}(U)\) and \(\bm{M}(U)\), inspired by the Dyson series. These general expressions are series solutions similar to the Dyson series. Our results show that the velocity memory effect emerges starting from the second-order correction. While most studies focus on linear polarization for simplicity, we compute the memory effect considering both polarizations.
Although the memory effect calculated here is nonlinear with respect to the gravitational wave amplitude, the velocity memory effect obtained from the nonlinear expansion of a yields \(\bm{P}\), resulting in a conclusion that is entirely different from that of the linear expansion.

The structure of this paper is as follows. In Section \ref{sec:pw}, we first review the metric form of plane gravitational waves in Brinkmann coordinates and the corresponding geodesic motion. The memory effect is observed by measuring the angular deviation between two freely falling gyroscopes. In Section \ref{sec:gwm}, we present the iterative expansion of \(\bm{P}(U)\) and \(\bm{M}(U)\) using the Dyson series, extending to arbitrary orders and apply a toy model of gravitational collapse to analyze the memory effect. We find that no final angular deviation occurs at first order, regardless of whether the initial separation is in displacement or velocity along the \(X^2\) direction. It is only upon considering second-order corrections that the velocity memory effect emerges. In Section \ref{sec:xp}, we solve for the final angular deviation by considering both polarizations and provide an estimation for a binary system. In the final section, we present our conclusions.
	
	\section{Plane gravitational wave}\label{sec:pw}
	
	\subsection{Brinkmann coordinates}
	The theory of exact plane gravitational waves differs fundamentally from that of linear gravitational waves. For exact plane waves, the derivation relies on identifying geometries that share the same symmetries as plane waves in flat spacetime, independent of Einstein's equations \cite{harte2015optics}. The resulting metric, commonly expressed in Brinkmann coordinates \((U, V, X^2, X^3)\), is as follows:
	\begin{equation}
		g=2dUdV+\delta _{ij}dX^idX^j+DdU^2 \,,  \label{eq:g}
	\end{equation}
where $D$ is a scalar function of coordinates ($U,X^1,X^2$) with the form of 

	\begin{equation}
	D = K_{ij}X^iX^j=\frac{1}{2}A_+(U)\bigg[ (X^2)^2-(X^3)^2\bigg] +A_\times (U)X^2X^3 \,.
\end{equation}
Here 	\(K_{ij}\) is a traceless, symmetric \(2\times2\) matrix that describes exact plane waves in vacuum general relativity.
When the metric describe a solution to Einstein equations, it reduces to $tr(K) = 0$.

The line element in the local Lorentz frame \((\hat{t},\hat{x},\hat{y},\hat{z})\) is given by
\begin{equation}
	ds^2=\left(-1+\frac{\Phi}{2}\right)d\hat{t}^2+d\bar{x}^2+d\hat{y}^2+\left(1+\frac{\Phi}{2}\right)d\hat{z}^2+ \Phi d\hat{t}d\hat{z}\,,
\end{equation}
where we have introduced the gravitational wave potential $\Phi(\hat t + \hat z, \hat{x},\hat{y})\equiv\frac{1}{2}\ddot{h}_+(\hat{x}^2-\hat{y}^2)+\ddot{h}_{\times}\hat{x}\hat{y}$.
The transformation between the local Lorentz coordinates and the Brinkmann coordinates is given by\cite{divakarla2021first}
\begin{eqnarray*}
	U &=& \frac{1}{\sqrt{2}}\left(\hat{t} + \hat{z}\right) \,, \\
	V &= & \frac{1}{\sqrt{2}}\left(-\hat{t} + \hat{z}\right)\,,\\
	X^2 &=& \hat{x} \,,\quad X^3 = \hat{y}\,,
\end{eqnarray*}
and
\begin{eqnarray}\label{eq:aplus}
	A_{+,{\times}} (U) &=& \ddot{h}_{+,{\times}}\left(\hat{t} + \hat{z}\right)\,.
\end{eqnarray}
The locally Lorentz frame can approximate the gravitational wave effects described by Brinkmann coordinates. Our analysis proceeds in local Lorentz coordinates, in line with Refs.\cite{divakarla2021first, wang2023spin}. The transformation between these two coordinate systems, as defined in Eq.\eqref{eq:aplus}, 
is crucial for the calculations.

	\subsection{Geodesic equation}
	
	In this section, we review the more concise general solution to the geodesic equation provided by \cite{wang2023spin}, building upon the results of \cite{flanagan2020persistent}.
	We consider an geodesic with tangent vector \(\bm{u}\),the geodesic equation for coordinate \(U\) is
	\begin{equation}
		\begin{aligned}
			\frac{\mathrm{d^2} U}{\mathrm{d} \tau ^2} =0\,,
		\end{aligned} \label{eq:g1}
	\end{equation}
	Given the initial position \(U(\tau_0)=U_0\),the solution to Eq.\eqref{eq:g1} is given by:
	\begin{equation}
		\begin{aligned}
			U=\gamma(\tau-\tau_0)+U_0\,.
		\end{aligned} \label{eq:g4}
	\end{equation}
	The observer proper velocity reads \(\bm{u}=\frac{\mathrm{d} x^\mu }{\mathrm{d} \tau }\partial_\mu =\gamma(\partial _U+v^1\partial_V+v^a\partial_{X^a})\)
	in Brinkmann cooedinates,where the 'time dilation factor' $\gamma$ defined as
	\begin{equation}
		\begin{aligned}
			\gamma= \frac{\mathrm{d} U}{\mathrm{d} \tau }=\frac{1}{\sqrt{-2v^1-D-v^av_a}} \,,
		\end{aligned}
	\end{equation}
normalizes \(\bm{u}\) with respect to the plane wave metric \eqref{eq:g}, i.e. $g_{\mu\nu}u^\mu u^\nu = -1 $.
	Since there is a linear relationship between \(U\) and \(\tau\),we will use the coordinate \(U\) instead of the geodesic affine parameter \(\tau\) in the rest paper.
	Then, the geodesic equation of the other three coordinates is as follows  \cite{wang2023spin}:
	\begin{equation}
		\begin{aligned}
			\bm{\ddot{X}}=\bm{K}(U)\bm{X}\,, 
		\end{aligned} \label{eq:g2}
	\end{equation}
	\begin{equation}
		\begin{aligned}
			\ddot{V}=-\frac{1}{2}\bm{\dot{D}}-2\bm{\dot{X}}^T\bm{KX}\,.
		\end{aligned}  \label{eq:g3}
	\end{equation}
Hereafter the dot denotes the derivative with respect to $U$.
	With the initial position \(x^\mu(U_0)=(U_0,V_0,\bm{X_0})\) and the initial four-velocity \(u^\mu(U_0)=\gamma(1,\dot{V_0},\bm{\dot{X}_0})\), the general solutions to Eqs.\,\eqref{eq:g2} and \eqref{eq:g3} are given by:
	\begin{equation}
		\begin{aligned}
			\bm{X}=\bm{P}(U)\bm{X}_0+(U-U_0)\bm{H(U)}\bm{\dot{X}}_0\,,
		\end{aligned} \label{eq:g5}
	\end{equation}
	\begin{equation}
		\begin{aligned}
			V=V_0-\frac{1}{2}\bigg[\bm{X}^T\bm{\dot{X}}-\bm{X}_0^T\bm{\dot{X}}_0+\frac{1}{\gamma^2}(U-U_0)\bigg]\,,
		\end{aligned}  \label{eq:g6}
	\end{equation}
	and \begin{equation}
		\begin{aligned}
			\bm{\dot{X}}=\bm{\dot{P}}\bm{X}_0+   \bigg[ (U-U_0)\bm{\dot{H}}(U)+\bm{H}(U)\bigg] \bm{\dot{X}}_0\,.
		\end{aligned}  \label{eq:g9}
	\end{equation}
	Both \(\bm{P}\) and \(\bm{H}\) are \(2\times2\) matrices that satisfy the following equations, respectively:
	\begin{equation}
		\begin{aligned}
			\bm{\ddot{P}}=\bm{K}\bm{P}\,,
		\end{aligned}  \label{eq:g7}
	\end{equation}
	\begin{equation}
		\begin{aligned}
			(U-U_0)\bm{\ddot{H}}+2\bm{\dot{H}}=(U-U_0)\bm{K}\bm{H}\,.
		\end{aligned}  \label{eq:g0}
	\end{equation}

At this point, the initial conditions for Eqs.\eqref{eq:g7} and \eqref{eq:g0} are given by \(\bm{P}(U_0) = \bm{I}\), \(\bm{H}(U_0) = \bm{I}\), and \(\bm{\dot{P}}(U_0) = 0\), \(\bm{\dot{H}}(U_0) = 0\), respectively. 
If we define \((U - U_0)\bm{H} = \bm{M}\), then Eq. \eqref{eq:g0} becomes \(\bm{\ddot{M}} = \bm{K}\bm{M}\), with initial conditions \(\bm{M}(U_0) = 0\) and \(\bm{\dot{M}}(U_0) = \bm{I}\). 
It is evident that obtaining the nonlinear expansion terms for \(\bm{P}\) and \(\bm{M}\) depends on solving Eqs.\eqref{eq:g7} and \eqref{eq:g0}, for which the initial conditions are known. These equations can be solved iteratively using the Dyson series to derive the general forms of \(\bm{P}(U)\) and \(\bm{M}(U)\), as discussed in Section III.
In \cite{zhang2018sturm}, Sturm-Liouville theory is identified as key to the memory effect. However, for plane waves, it may be more direct to assert that the Dyson series plays a central role in the memory effect.

	\subsection{Precession of the gyroscope}
	In \cite{wang2023spin}, a method for observing the velocity memory effect statically in a plane wave spacetime is presented by comparing the angular deviation between two freely falling gyroscopes after passing through a gravitational wave region. One gyroscope, \(G_0\), is placed at the origin, while the other, \(G_1\), is positioned along the \(X^2\) axis. Initially, the gyroscopes are aligned, and after passing through the gravitational wave region, the angular deviation between the two is measured.
	It is shown that the precession angle is independent of the specific path, allowing the gyroscopes' worldlines to overlap for a direct comparison of their angular deviation. A freely falling gyroscope has a spin vector \(S\) that obeys Fermi-Walker transport:
	\[
	(\bm{u} \cdot \nabla \bm{S})^\mu = (u^\mu a^\nu - u^\nu a^\mu) S_\nu,
	\]
	where \(\bm{e}_{\hat{\mu}}\) is a local tetrad at the gyroscope’s location, with \(\bm{e}_{\hat{0}} = \bm{u}\) and \(\bm{e}_{\hat{\mu}} \cdot \bm{e}_{\hat{\nu}} = \eta_{\hat{\mu} \hat{\nu}}\). This allows the parallel transport to be expressed as the precession equation.
	\begin{equation}
		\begin{aligned}
			\frac{\mathrm{d} S^{\hat{i}}}{\mathrm{d} \tau } =\Omega ^{\hat{i}}_{\hat{j}}S^{\hat{j}}=-u^a\omega _a^{\hat{i}\hat{j}}S^{\hat{j}}\,,
		\end{aligned}
	\end{equation} 
	when \(\omega _a^{\hat{i}{j}}\) is the spin connection one-form associated with the tetrad \({\bm{e}_{\hat{\mu}}}\),and \(\Omega\) canbe considered as a 2-form of angular velocity.
	The resulting angular deviation between the two gyroscopes is thus given by  \cite{wang2023spin} :
	\begin{equation}
		\begin{aligned}
			\Theta^a=-\arcsin\left(\frac{v^a}{\sqrt{2}}\right)\,.
		\end{aligned}  \label{eq:g8}
	\end{equation}
The angular deviation between two gyroscopes depends on \(v\), and thus, it is essentially a manifestation of the velocity memory effect. The angular deviation is calculated under initial conditions of a separation displacement \(L\) along the \(X^2\)-direction and a separation velocity \(v_0\). To facilitate the observation of this angular deviation, numerical simulations are performed within the gravitational collapse model, where both \(\bm{P}\) and \(\bm{M}\) are expanded to linear order.
The TT-gauge metric is a trivial linearization of the Baldwin-Jeffery-Rosen (BJR) coordinates \cite{harte2015optics}. The linear expansion of \(P\) utilizes the relationship between Brinkmann coordinates and BJR coordinates \cite{zhang2017soft}. From \(a_{ij} = \delta_{ij} + h_{ij}(u) + \dots\), the expression \(P_{ij} = \delta_{ij} + \frac{1}{2} h_{ij}(u) + \dots\) is obtained. The BJR coordinates allow for a comparison between linear and nonlinear gravitational waves in the TT gauge.
In Section \ref{sec:gwm}, it will be shown that, without resorting to the B coordinates, the expressions for \(\bm{P}\) and \(\bm{M}\) can be derived by analogy with the Dyson series.

	\section{Gravitational wave velocity memory effect}\label{sec:gwm}
	
	\subsection{Iteration of Dyson series}
	
	The second-order differential equation \(\bm{\ddot{P}}(U) = \bm{K}(U)\bm{P}(U)\), 
	along with the initial conditions \( \bm{P}(U_0) = \bm{I}, \bm{\dot{P}}(U_0) = 0 \),
	allows \[
	\bm{P}(U)=\bm{I}+\int_{U_0}^{U}dU_1\int_{U_0}^{U_1}dU_2\bm{K}(U_2)\bm{P}(U_2)
	\] to be expressed. 
	By iteratively applying \(\bm{P(U)}\), the general solution is obtained,namely\cite{harte2015optics}
\begin{equation}	\label{eq:PU}
	\bm{P}(U)=\bm{I}+ \sum_{j=1}^{n}\left(\int_{U_0}^{U}dU_{1}\int_{U_0}^{U_{1}}dU_{2}\bm{K}(U_{2})\prod_{i=2      }^{j}\int_{U_0}^{U_{2i-2}}dU_{2i-1}\int_{U_0}^{U_{2i-1}}dU_{2i}\bm{K}(U_{2i}) \right)\,,
\end{equation}
	Let \((U-U_0)\bm{H}(U)=\bm{M}(U)\) hold, then \(\bm{\dot{M}}(U)=(U-U_0)\bm{\dot{H}}(U)+\bm{H}(U)\),\(\bm{\ddot{M}}(U)=\bm{K}(U)\bm{M}(U)\) follows. 
	With the initial conditions given by \(\bm{M}(U_0)=0\),\(\bm{\dot{M}}(U_0)=\bm{I}\), the result is then obtained,
	\[
	\bm{M}(U)=\int_{U_0}^{U}dU_1\bm{I}+\int_{U_0}^{U}dU_1\int_{U_0}^{U_1}dU_2\bm{K}(U_2) \bm{M}(U_2)\,.
	\]
	By iteratively applying \(\bm{M(U)}\), the final result is obtained,
	\begin{equation}
	\bm{M}(U)=\int_{U_0}^{U}dU_1\bm{I}+ \sum^{n}_{j=1}\left(\int_{U_0}^{U}dU_{1}\int_{U_0}^{U_{1}}dU_{2}\bm{K}(U_{2})  \prod^{j}_{i=2} \int_{U_0}^{U_{2i-2}}dU_{2i-1}\int_{U_0}^{U_{2i-1}}dU_{2i}\bm{K}(U_{2i})\int_{U_0}^{U_{2j}} dU_{2j+1} \bm{I}\right)\,.
	\label{eq:MU}
\end{equation}
From Eqs.\eqref{eq:PU} and \eqref{eq:MU} can be expanded to arbitrary orders. Note that \(\tau\) is the time-ordering operator.

	\subsection{Separation Displacement}
	When only the initial conditions are considered, with a separation displacement \(L\) along the \(X^2\)-direction, \(\Delta V = \dot{P}L\) can be obtained from the initial conditions \(X_0 = L\), \(\dot{X}_0 = 0\), and Eq.\eqref{eq:g9}. Using Eq.\eqref{eq:g8}, the angular deviation is determined. For clarity, its matrix form is explicitly given as follows:
	\begin{equation}
		\begin{pmatrix}
			\Delta \Theta ^2\\
			\Delta \Theta^3 
		\end{pmatrix}=-\arcsin\left (\frac{1}{\sqrt{2} } \begin{pmatrix}
			\dot{P}_{22}  & \dot{P}_{23}  \\
			\dot{P}_{32}  & \dot{P}_{33} 
		\end{pmatrix}\begin{pmatrix}
			L\\
			0
		\end{pmatrix}\right)=-\arcsin\left(\frac{1}{\sqrt{2} }\begin{pmatrix}
			\dot{P}_{22}L  \\
			\dot{P}_{32}L 
		\end{pmatrix} \right)\,.  \label{eq:L1} 
	\end{equation}
	To determine the angle difference, one must first calculate \({P}_{22}\) and \({P}_{32}\), after which the matrix expansion of \({P}\) can be written as 

\begin{eqnarray}
&&		\begin{pmatrix}
		P_{22}(U) & P_{23}(U) \\
		P_{32}(U) & P_{33}(U)
	\end{pmatrix}
=  \begin{pmatrix}
		1 & 0 \\
		0 & 1
	\end{pmatrix} 
	+ \frac{1}{2}	 \int_{U_0}^{U} dU_1 \int_{U_0}^{U_1} dU_2
\begin{pmatrix}
		A_+(U_2) & A_\times(U_2) \\
	A_\times(U_2) & -A_+(U_2)
	\end{pmatrix} \nonumber \\
	&+&\frac{1}{4}	 \int_{U_0}^{U} dU_1 \int_{U_0}^{U_1} dU_2
	\begin{pmatrix}
		A_+(U_2) & A_\times(U_2) \\
	A_\times(U_2) & -A_+(U_2)
	\end{pmatrix}
	\int_{U_0}^{U_2} dU_3 \int_{U_0}^{U_3} dU_4
	\begin{pmatrix}
	A_+(U_4) & A_\times(U_4) \\
	A_\times(U_4) & -A_+(U_4)
	\end{pmatrix} + \cdots\,.  \label{eq:L2}
\end{eqnarray}

\begin{figure}[h!]
		\centering
		\includegraphics[width=0.8\textwidth]{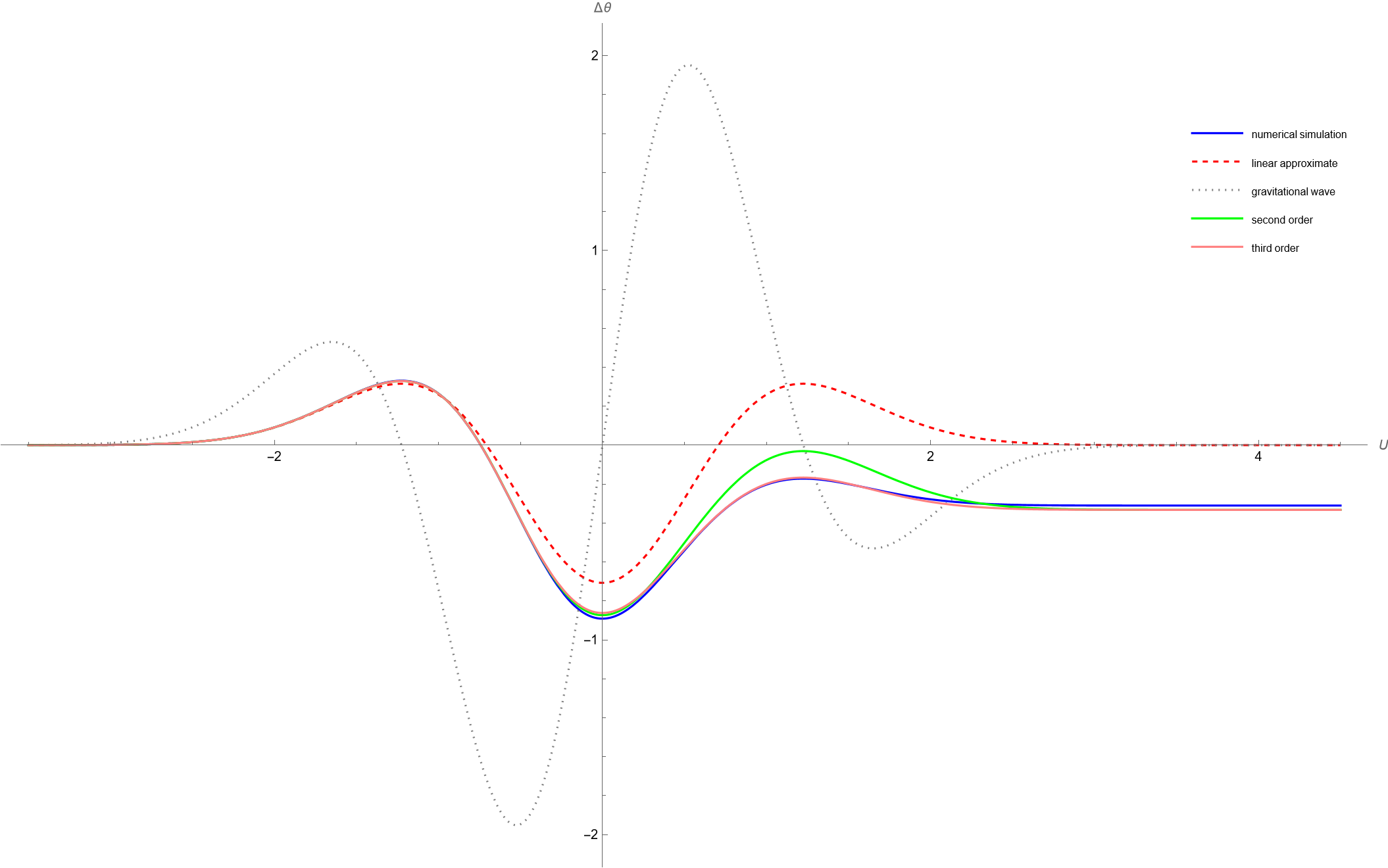}
		\caption{
			The angular deviation between two gyroscopes is analyzed as it evolves with initial separation displacement. For clarity in the figures, a negative sign is applied to all angle differences, allowing focus on the relative deviation without affecting the results. \label{fig:mathematica_plot18}}
	\end{figure}

To facilitate the observation of angular variations, we use the toy model of gravitational collapse from \cite{zhang2017memory}, namely:
	\begin{equation}
		\begin{aligned}
			A_+(U)=\ddot{h}_+=\frac{1}{2}\frac{\mathrm{d}^3 (e^{-U^2})}{\mathrm{d} U^3}\,.
		\end{aligned}   \label{eq:L3}
	\end{equation}
	For simplicity, we begin by considering linear polarization, where \(A_\times=0\). Later, the case of 
	\(A_\times\neq0\) will also be examined. By considering a third-order expansion and substituting \eqref{eq:L3} into \eqref{eq:L2},we obtain:
	\begin{equation}
	\begin{aligned}
		P_{22} &= 1 + \frac{1}{2}h_+(U) + \frac{1}{8}h_+(U)^2 + \frac{1}{48}h_+(U)^3 \\
		&  - \frac{1}{4}\int_{U_0}^{U}dU_1\int_{U_0}^{U_1}dU_2\dot{h}_+(U_2)^2 
		  - \frac{1}{8}\int_{U_0}^{U}dU_1\int_{U_0}^{U_1}dU_2\dot{h}_+(U_2)^2h_+(U_2) \\
		& - \frac{1}{8} \int_{U_0}^{U}dU_1\int_{U_0}^{U_1}dU_2\ddot{h}_+(U_2)\int_{U_0}^{U_2}dU_3\int_{U_0}^{U_3}dU_4\dot{h}_+(U_4)^2 + \dots\,.
	\end{aligned}   
\end{equation}
	So that
	\begin{equation}
		\begin{aligned}
			\Delta \Theta^2 &= -\arcsin\left(\frac{1}{\sqrt{2}}\dot{P}_{22}L\right) \\
			&= -\frac{L}{\sqrt{2}}\bigg(\frac{1}{2}\dot{h}_+(U) + \frac{1}{4}\dot{h}_+(U)h_+(U) 
			+ \frac{1}{16}\dot{h}_+(U)h_+(U)^2 	- \frac{1}{4}\int_{U_0}^{U} dU_1 \dot{h}_+(U_1)^2  \\
			& - \frac{1}{8}\int_{U_0}^{U} dU_1 \dot{h}_+(U_1)^2 h_+(U_1)
			- \frac{1}{8} \int_{U_0}^{U} dU_1 \ddot{h}_+(U_1) \int_{U_0}^{U_1} dU_2 \int_{U_0}^{U_2} dU_3 \dot{h}_+(U_3)^2 + \dots \bigg)\,,
		\end{aligned}  \label{eq:p22}
	\end{equation}
for  small angles. From Fig.\ref{fig:mathematica_plot18}, it is clear that when considering only the initial conditions with a separation displacement \(L\) along the \(X^2\) direction, no final angular deviation arises in the linear approximation of the simplified gravitational collapse model. However, an angular deviation emerges starting from the second-order approximation, confirming that higher-order terms introduce angular deviation \cite{wang2023spin}, beginning at second order. By the third-order approximation, the results closely match numerical simulations. In our calculations, this assumption is implicitly applied for simplicity, specifically regarding \(h_+(U_0) = 0\) and \(\dot{h}_+(U_0) = 0\).
	
	\subsection{Separation Velocity}
\begin{figure}[h!]
		\centering
		\includegraphics[width=0.8\textwidth]{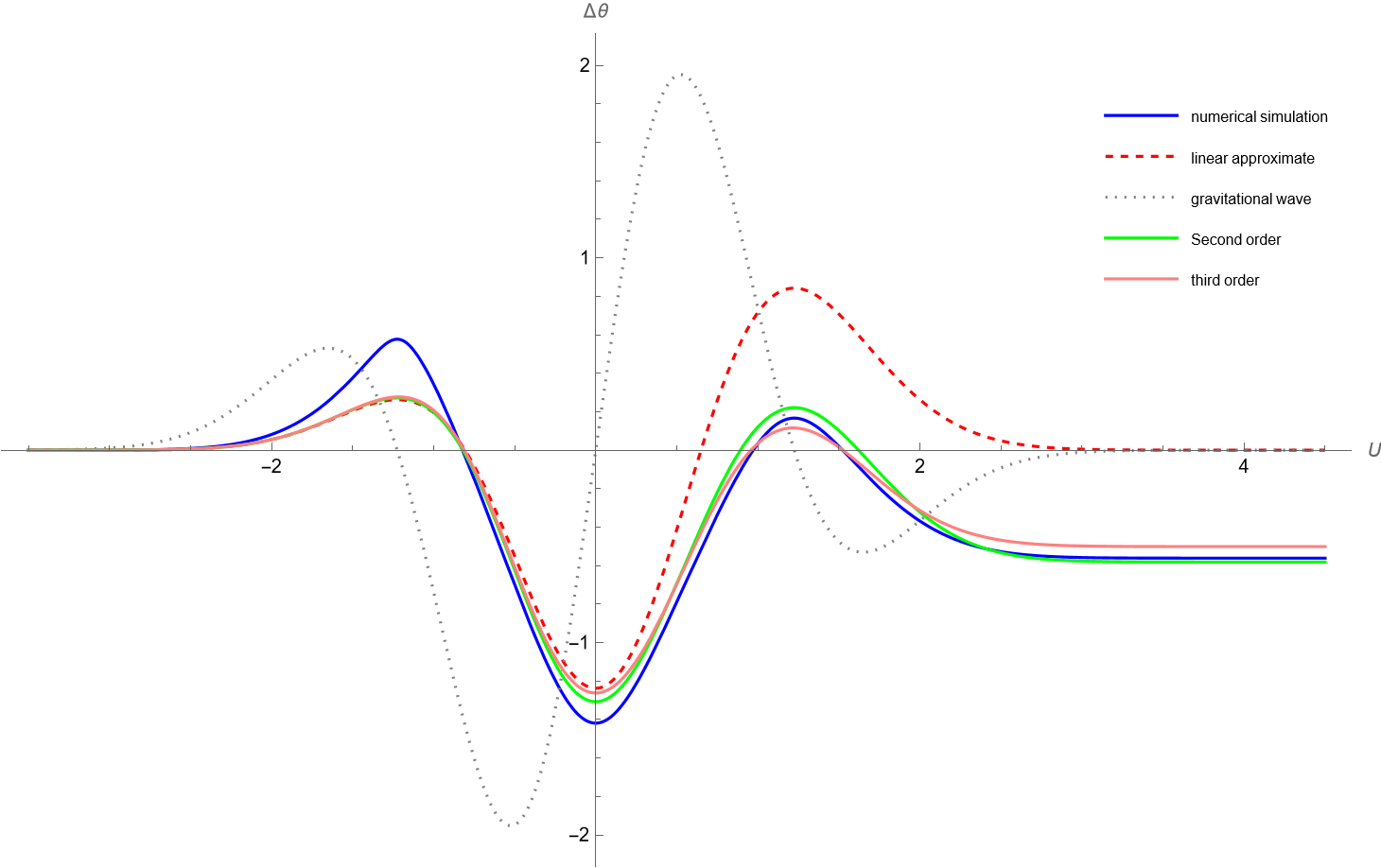}
		\caption{
			The angular deviation between two gyroscopes evolves as a function of the initial separation velocity. \label{fig:mathematica_plog}}
	\end{figure}

	Considering only the initial conditions, with separation velocity along the \(X^2\) direction at a rate of \(V_0\), the initial state is defined as \(\dot{X}_0=V_0\).
	To provide a clear representation of the angular deviation, we begin by expressing it in matrix form:
	\begin{equation}   
		\begin{pmatrix}
			\Delta \Theta^2  \\
			\Delta \Theta ^3
		\end{pmatrix}=-\bigg[\arcsin(\frac{1}{\sqrt{2} } \begin{pmatrix}
			V^2 \\
			V^3
		\end{pmatrix})-\arcsin(\frac{1}{\sqrt{2}}\begin{pmatrix}
			V_0\\
			0
		\end{pmatrix})\bigg]\,, \label{eq:V}
	\end{equation}
	where \(V^2\) and \(V^3\) must first be calculated 	to determine the angular deviation:
	\begin{equation}
		\begin{pmatrix}
			V^2\\
			V^3
		\end{pmatrix}=\bigg( (U-U_0)\begin{pmatrix}
			\dot{H}_{22}  & \dot{H}_{23} \\
			\dot{H}_{32 } & \dot{H}_{33} 
		\end{pmatrix}+\begin{pmatrix}
			H_{22} & H_{23}\\
			H_{32} & H_{33}
		\end{pmatrix} \bigg) 
	 \begin{pmatrix}
			V_0\\
			0
		\end{pmatrix}=\bigg ((U-U_0)\begin{pmatrix}
			\dot{H}_{22} \\
			\dot{H}_{32} 
		\end{pmatrix}+\begin{pmatrix}
			H_{22} \\
			H_{32}
		\end{pmatrix}\bigg)V_0 =\begin{pmatrix}
			\dot{M}_{22}\\ \dot{M}_{32}
			
		\end{pmatrix}V_0\,. \label{eq:V1}
	\end{equation}
At the beginning, \(M_{22}\) and \(M_{32}\) are first calculated. For simplicity, we initially focus on the \(+\) polarization, where \(M_{32} = 0\) and \(V^3 = 0\). Expanding \(M_{22}\) to third order yields \(V^2\) as follows:
\begin{equation}
	\begin{aligned}
		V^2 = &\left((U-U_0)\dot{H}_{22} + H_{22}\right)V_0 = \dot{M}_{22}V_0 \\
		=&\Bigg(1 - \frac{1}{2}h_+(U) + \frac{1}{2}\dot{h}_+(U)(U - U_0) 
		+ \frac{1}{4}\dot{h}_+(U)h_+(U)(U - U_0) - \frac{1}{8}h_+(U)^2 \\
		&- \frac{1}{4}\int_{U_0}^{U} dU_1 \dot{h}_+(U_1)^2(U_1 - U_0) 
		- \frac{1}{2}\int_{U_0}^{U} dU_1 \ddot{h}_+(U_1)\int_{U_0}^{U_1} dU_2 h_+(U_2) \\
		&+ \frac{1}{8}\dot{h}_+(U)h_+(U)^2U 
		- \frac{1}{4}\int_{U_0}^{U} dU_1 \dot{h}_+(U_1)^2U_1 
		- \frac{1}{24}h_+(U)^3 \\
		&- \frac{1}{8}\int_{U_0}^{U} dU_1 \ddot{h}_+(U_1)\int_{U_0}^{U_1} dU_2 \dot{h}_+(U_2)h_+(U_2)U_2 
		- \frac{3}{16}\int_{U_0}^{U} dU_1 \ddot{h}_+(U_1)\int_{U_0}^{U_1} dU_2 h_+(U_2)^2 \\
		&- \frac{1}{8}\int_{U_0}^{U} dU_1 \ddot{h}_+(U_1)\int_{U_0}^{U_1} dU_2 \int_{U_0}^{U_2} dU_3 \dot{h}_+(U_3)^2U_3 
		- \frac{1}{4}\int_{U_0}^{U} dU_1 \ddot{h}_+(U_1)\int_{U_0}^{U_1} dU_2 \int_{U_0}^{U_2} dU_3 \\
		&\ddot{h}_+(U_3)\int_{U_0}^{U_3} dU_4 h_+(U_4) 
		- \frac{1}{16}\dot{h}_+(U)h_+(U)^2U_0 
		+ \frac{1}{8}\int_{U_0}^{U} dU_1 \dot{h}_+(U_1)^2h_+(U_1)U_0 \\
		&+ \frac{1}{8}\int_{U_0}^{U} dU_1 \ddot{h}_+(U_1)\int_{U_0}^{U_1} dU_2 \int_{U_0}^{U_2} dU_3 \dot{h}_+(U_3)^2U_0\Bigg)V_0\,.
	\end{aligned}
\end{equation}
	Thus, we have \(\Delta \Theta^2\) :
	\begin{equation}
 \begin{aligned}
	\Delta \Theta^2 &=-\bigg[ \arcsin\left(\frac{V^2}{\sqrt{2}}\right) - \arcsin\left(\frac{V_0}{\sqrt{2}}\right) \bigg]\\
	&=- \frac{V_0}{\sqrt{2}} \Bigg(- \frac{1}{2}h_+(U) + \frac{1}{2}\dot{h}_+(U)(U - U_0) 
	+ \frac{1}{4}\dot{h}_+(U)h_+(U)(U - U_0) - \frac{1}{8}h_+(U)^2 \\
	&\quad - \frac{1}{4}\int_{U_0}^{U} dU_1 \dot{h}_+(U_1)^2(U_1 - U_0) 
	- \frac{1}{2}\int_{U_0}^{U} dU_1 \ddot{h}_+(U_1)\int_{U_0}^{U_1} dU_2 h_+(U_2) \\
	&\quad + \frac{1}{8}\dot{h}_+(U)h_+(U)^2U 
	- \frac{1}{4}\int_{U_0}^{U} dU_1 \dot{h}_+(U_1)^2U_1 
	- \frac{1}{24}h_+(U)^3 \\
	&\quad - \frac{1}{8}\int_{U_0}^{U} dU_1 \ddot{h}_+(U_1)\int_{U_0}^{U_1} dU_2 \dot{h}_+(U_2)h_+(U_2)U_2 
	- \frac{3}{16}\int_{U_0}^{U} dU_1 \ddot{h}_+(U_1)\int_{U_0}^{U_1} dU_2 h_+(U_2)^2 \\
	&\quad - \frac{1}{8}\int_{U_0}^{U} dU_1 \ddot{h}_+(U_1)\int_{U_0}^{U_1} dU_2 \int_{U_0}^{U_2} dU_3 \dot{h}_+(U_3)^2U_3 
	- \frac{1}{4}\int_{U_0}^{U} dU_1 \ddot{h}_+(U_1)\int_{U_0}^{U_1} dU_2 \int_{U_0}^{U_2} dU_3 \\
	&\quad \ddot{h}_+(U_3)\int_{U_0}^{U_3} dU_4 h_+(U_4) 
	- \frac{1}{16}\dot{h}_+(U)h_+(U)^2U_0 
	+ \frac{1}{8}\int_{U_0}^{U} dU_1 \dot{h}_+(U_1)^2h_+(U_1)U_0 \\
	&\quad + \frac{1}{8}\int_{U_0}^{U} dU_1 \ddot{h}_+(U_1)\int_{U_0}^{U_1} dU_2 \int_{U_0}^{U_2} dU_3 \dot{h}_+(U_3)^2U_0 \Bigg)\,.
\end{aligned}  \label{eq:M22}
\end{equation}
As shown in Fig.\ref{fig:mathematica_plog}, for initial conditions with a separation velocity \(V_0\) only in the \(X^2\) direction, no final angular deviation is observed under the linear approximation. However, when second-order and third-order corrections are included, a final angular deviation emerges.

	\subsection{The initial conditions include both separation displacement and separation velocity.}
	When the initial conditions are considered, with both displacement and velocity separated, the analysis is limited to second order. Using
	Eqs.\,\eqref{eq:g9},\eqref{eq:g8},\eqref{eq:p22} and \eqref{eq:M22},we have
	\begin{equation}
		\begin{aligned}
			\Delta \Theta^2 &=-\frac{1}{\sqrt{2}} (\dot{P}_{22}L+\dot{M}_{22}V_0) \\
			&=- \frac{L}{\sqrt{2}}\left(\frac{1}{2}\dot{h}_+(U) + \frac{1}{4}\dot{h}_+(U)h_+(U) 
			- \frac{1}{4}\int_{U_0}^{U} dU_1 \dot{h}_+(U_1)^2\right) \\
			&- \frac{V_0}{\sqrt{2}} \bigg(- \frac{1}{2}h_+(U) + \frac{1}{2}\dot{h}_+(U)(U - U_0) 
			+ \frac{1}{4}\dot{h}_+(U)h_+(U)(U - U_0) - \frac{1}{8}h_+(U)^2 \\
			&\quad - \frac{1}{4}\int_{U_0}^{U} dU_1 \dot{h}_+(U_1)^2(U_1 - U_0) 
			- \frac{1}{2}\int_{U_0}^{U} dU_1 \ddot{h}_+(U_1)\int_{U_0}^{U_1} dU_2 h_+(U_2) \bigg)\,. \\
		\end{aligned}  \label{eq:LV}
	\end{equation}
	The oscillation modes are inactive, and restore the standard unit, we can get
	\begin{equation}
		\begin{aligned}
			\Delta \Theta^2 &= \frac{L}{\sqrt{2}c}\bigg(  \frac{1}{4}\int_{U_0}^{U} dU_1 \dot{h}_+(U_1)^2\bigg) 
			- \frac{V_0}{\sqrt{2}c}\bigg (- \frac{1}{2}h_+(U) - \frac{1}{8}h_+(U)^2 \\
			&\quad - \frac{1}{4}\int_{U_0}^{U} dU_1 \dot{h}_+(U_1)^2(U_1 - U_0) 
			- \frac{1}{2}\int_{U_0}^{U} dU_1 \ddot{h}_+(U_1)\int_{U_0}^{U_1} dU_2 h_+(U_2) \bigg) \,. \\
		\end{aligned}
	\end{equation}
For the plane wave approximation to hold, the maximum initial separation distance \(L\) between two gyroscopes with a separation velocity \(v_0 \ll c\) is \(10^{17}\) meters \cite{divakarla2021first,wang2023spin}. Considering that the strain amplitude from the merger of two supermassive black holes is about \(10^{-15}\) \cite{aggarwal2019nanograv}, the resulting contribution angle due to the initial separation \(L\) is approximately \(10^{-23}\) radians. The contribution from the initial separation displacement only becomes significant at higher orders \cite{wang2023spin}, specifically at second order. This is because the first-order expansion of \(P = 1 + \frac{1}{2}h\) with respect to the initial separation \(L\) corresponds to the linear gravitational waves, which do not generate a velocity memory effect. The angular formula \(\theta = -\frac{L}{2\sqrt{2}} \dot{h}_+(U)\), derived from the initial separation distance \(L\), corresponds to the first-order expansion \(P = 1 + \frac{1}{2}h\) of \(\bm{P}\), where the oscillating term does not contribute to the memory effect.

	\section{CONSIDER X POLARIZATION} \label{sec:xp}
	
Most studies simplify the analysis by considering only a single polarization,
	reducing \eqref{eq:g2} to two independent second-order differential equations. However, including both polarizations results in two coupled second-order differential equations.
	Here, we present the resulting velocity memory effect when both polarizations are taken into account.
	Incorporating the \({\times}\) polarization allows calculations to arbitrary orders for \(M(U)\) and \(P(U)\), here, we limit the expansion to second order for clarity. 
	Using \eqref{eq:L1} and \eqref{eq:L2},we derive the following expressions:
	\begin{eqnarray}
	P_{22}(U) &=& 1 + \frac{1}{2}h_+(U) + \frac{1}{8}h_+(U)^2 - \frac{1}{4}\int_{U_0}^{U}dU_1\int_{U_0}^{U_1}dU_2\dot{h}_+(U_2)^2 + \frac{1}{8}h_\times(U)^2-\frac{1}{4}\int_{U_0}^{U}dU_1\int_{U_0}^{U_1}dU_2\dot{h}_\times(U_2)^2+\dots\,, \\
P_{32}(U) &=&  \frac{1}{2} h_\times (U) 
+ \frac{1}{4} \int_{U_0}^{U} dU_1 \int_{U_0}^{U_1} dU_2 \ddot{h}_\times(U_2) h_+ (U_2) - \frac{1}{4} \int_{U_0}^{U} dU_1 \int_{U_0}^{U_1} dU_2 \ddot{h}_+ (U_2) h_\times(U_2) + \dots\,,		
	\end{eqnarray}
and
\begin{eqnarray}
	\nonumber
	\Delta \Theta^2_1 &=& -\arcsin\left(\frac{1}{\sqrt{2}}\dot{P}_{22}L\right) \\
	&=& -\frac{L}{\sqrt{2}}\left(\frac{1}{2}\dot{h}_+(U) + \frac{1}{4}h_+(U)\dot{h}_+(U) - \frac{1}{4}\int_{U_0}^{U}dU_1\dot{h}_+(U_1)^2 + \frac{1}{4}h_\times (U)\dot{h}_\times (U) - \frac{1}{4}\int_{U_0}^{U}dU_1\dot{h}_\times (U_1)^2 \right)\,, \\
	\nonumber
				\Delta \Theta^3_1 &=& -\arcsin\left(\frac{1}{\sqrt{2}}\dot{P}_{32}L\right) \\
	&=& -\frac{L}{\sqrt{2}}\left( \frac{1}{2}\dot{h}_{\times}(U) 
	+ \frac{1}{4}\int_{U_0}^{U} dU_1 \ddot{h}_{\times} (U_1)h_+(U_1) 
	- \frac{1}{4}\int_{U_0}^{U} dU_1 \ddot{h}_+(U_1) h_{\times} (U_1)  \right) + \dots\,.
\end{eqnarray}
	\begin{figure}[h!]
		\centering
		\includegraphics[width=0.8\textwidth]{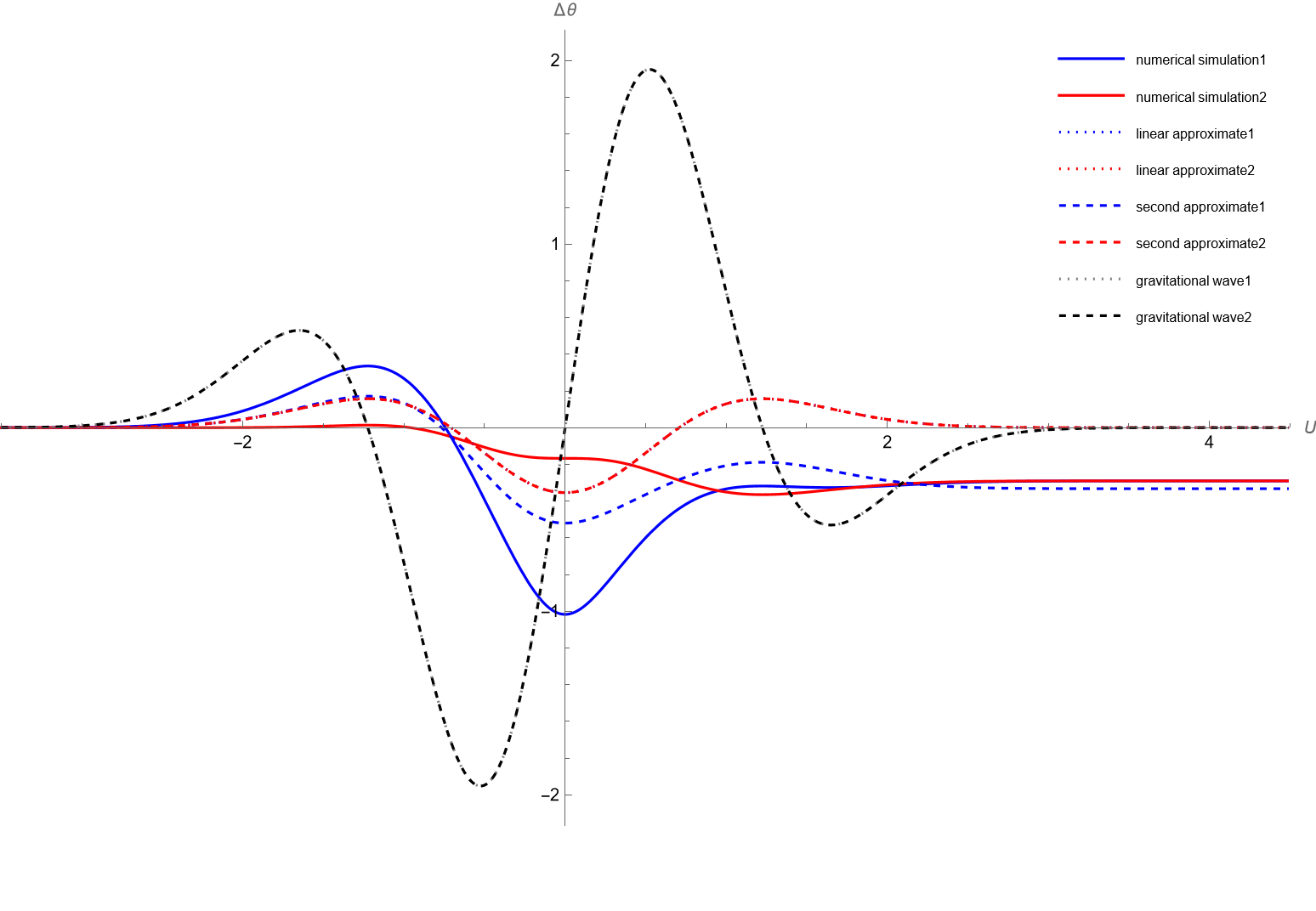}
		\caption{The angular deviation between two gyroscopes evolves with the initial separation displacement, taking into account the x polarization. \label{fig:mathematica_plot15}}
	\end{figure}
	
\begin{figure}[h!]
		\centering
		\includegraphics[width=0.8\textwidth]{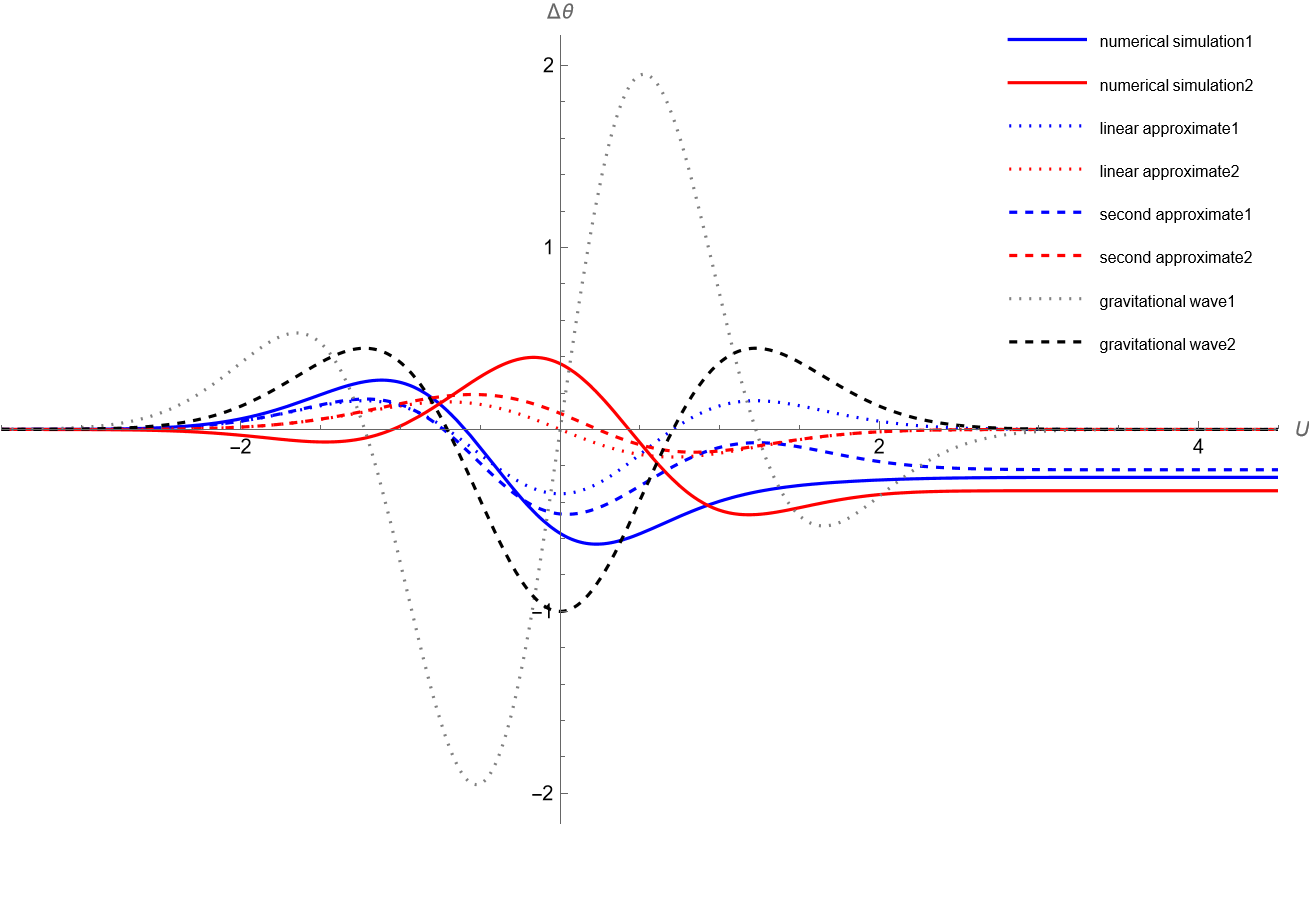}
		\caption{The angular deviation between two gyroscopes evolves with initial separation displacement under the influence of x polarization. In this analysis,the + polarization is applied to the gravitational collapse model \eqref{eq:L3},while the X polarization is considered in \eqref{eq:A_x}. \label{fig:mathematica_plot16}}
	\end{figure}

	When both polarizations are included in gravitational collapse model \eqref{eq:L3}, the results are shown in Fig.\ref{fig:mathematica_plot15}.
	It is observed that, with an initial separation displacement in the \(X^2\) direction, neither the linear approximation nor the second-order correction of \(P_{32}\) produces a memory effect; memory effects only emerge in higher-order corrections.
	Although this approach aids in simplifying the analysis, actual gravitational waves are inherently more complex. Notably, the angular deviation appears identical in both directions. 
	To eliminate the possibility that these findings are specific to a single model, we further investigate the gravitational collapse model with \(+\) polarization and consider an alternative model under \(\times\) polarization, as described below:
	\begin{equation}
		\begin{aligned}
			A_\times(U)=\frac{1}{2}\frac{\mathrm{d}^2 (e^{-U^2})}{\mathrm{d} U^2}=\ddot{h}_\times+o(h^2)\,.
		\end{aligned}   \label{eq:A_x}
	\end{equation}
	The results, displayed in Fig.\ref{fig:mathematica_plot16}, align with those obtained from single-model analyses. Additionally, third-order corrections for \(P_{22}\) and \(P_{32}\) were examined, revealing no final angular deviation in the third-order correction for \(P_{32}\).
	Although higher-order corrections could be computed, these calculations become complex due to the need for multiple matrix multiplications. Therefore, we do not extend the calculations here to determine the precise order at which a final angular deviation might appear.
	In this setup, gyroscope \(G_0\) is placed at the origin of the \(X^2\)-axis.  while gyroscope \(G_1\)is positioned along the \(X^2\)-axis. 
	To facilitate angle measurements, gyroscope \(G_0^{'}\) is located at the origin of the \(X^3\) and aligned with the \(X^3\)-axis. Consequently,\(\theta^2\) represents the angle between the gyroscopes at the origins of the \(G_1\) and \(X^2\),while \(\theta^3\) denotes the angle relative to the gyroscope along the \(X^3\)-axis.
	
	Considering the initial conditions as separation velocity, we obtain the following results:
	\begin{equation}
		\begin{aligned}
			\Delta \Theta ^2_2= & -\bigg[\arcsin(\frac{V^2}{\sqrt{2}})-\arcsin(\frac{V_0}{\sqrt{2}}) \bigg]\\
			= & -\frac{V_0}{\sqrt{2}}\bigg[-\frac{1}{2}h_+(U)+\frac{1}{2}\dot{h}_+(U)(U-U_0)\\
			&+\frac{1}{4}\dot{h}_+(U)h_+(U)(U-U_0)-\frac{1}{8}h_+(U)^2-\frac{1}{4}\int_{U_0}^{U}dU_1\dot{h}_+(U_1)^2(U_1-U_0)-\frac{1}{2}\int_{U_0}^{U}dU_1\ddot{h}_+(U_1)\int_{U_0}^{U_1}dU_2h_+(U_2) \\
			&+\frac{1}{4}\dot{h}_\times (U)h_\times (U)(U-U_0)-\frac{1}{8}h_\times (U)^2-\frac{1}{4}\int_{U_0}^{U}dU_1\dot{h}_\times (U_1)^2(U-U_0)-\frac{1}{2}\int_{U_0}^{U}dU_1\ddot{h}_\times (U_1)\int_{U_0}^{U_1}dU_2h_\times (U_2)\bigg]\,,
		\end{aligned}
	\end{equation}
	\begin{equation}
		\begin{aligned}
			\Delta \Theta ^3_2= &-\bigg[ \arcsin(\frac{V^3}{\sqrt{2}})-\arcsin(\frac{V_0}{\sqrt{2}}) \bigg]\\
			= & -\frac{V_0}{\sqrt{2}}\bigg[-1-\frac{1}{2}h_\times(U)+\frac{1}{2}\dot{h}_\times(U)(U-U_0) \\
			& +\frac{1}{4}\int_{U_0}^{U}dU_1\ddot{h}_\times(U_1)h_+(U_1)U_1 -\frac{1}{2}\int_{U_0}^{U}dU_1\ddot{h}_\times (U_1)\int_{U_0}^{U_1}dU_2h_+(U_2)-\frac{1}{4}\int_{U_0}^{U}dU_1 \ddot{h}_\times (U_1)h_+(U_1)U_0 \\
			& -(\frac{1}{4}\int_{U_0}^{U}dU_1\ddot{h}_+(U_1)h_\times (U_1)U_1 -\frac{1}{2}\int_{U_0}^{U}dU_1\ddot{h}_+(U_1)\int_{U_0}^{U_1}dU_2h_\times (U_2)-\frac{1}{4}\int_{U_0}^{U}dU_1 \ddot{h}_+ (U_1)h_\times (U_1)U_0)\bigg]\,.
		\end{aligned}
	\end{equation}
	Here \(\Delta \Theta^2\) denotes the angular deviation of \(G_1\) relative to \(G_0\),while \(\Delta \Theta^3\) represents the angular difference of \(G_1\) relative to \(G_0^{'}\).
	The influence of \(\times\) polarization is examined, which introduces \(\theta_3\).Oscillatory terms do not contribute to the memory effect. When initial conditions include both separation displacement and velocity along the \(X^2\) direction, the final angular deviation is given by:
	\begin{equation}
		\begin{aligned}
			\Delta \Theta^2 &=-\frac{L}{\sqrt{2}}\left(- \frac{1}{4}\int_{U_0}^{U}dU_1\dot{h}_+(U_1)^2  - \frac{1}{4}\int_{U_0}^{U}dU_1\dot{h}_\times (U_1)^2 \right)\\
			& -\frac{V_0}{\sqrt{2}}\bigg[-\frac{1}{2}h_+(U)\\
			&-\frac{1}{8}h_+(U)^2-\frac{1}{4}\int_{U_0}^{U}dU_1\dot{h}_+(U_1)^2(U_1-U_0)-\frac{1}{2}\int_{U_0}^{U}dU_1\ddot{h}_+(U_1)\int_{U_0}^{U_1}dU_2h_+(U_2) \\
			&-\frac{1}{8}h_\times (U)^2-\frac{1}{4}\int_{U_0}^{U}dU_1\dot{h}_\times (U_1)^2(U-U_0)-\frac{1}{2}\int_{U_0}^{U}dU_1\ddot{h}_\times (U_1)\int_{U_0}^{U_1}dU_2h_\times (U_2)\bigg]\,,
		\end{aligned}
	\end{equation}
	
	\begin{equation}
		\begin{aligned}
			\Delta \Theta^3 &=-\frac{L}{\sqrt{2}}\left( 
			\frac{1}{4}\int_{U_0}^{U} dU_1 \ddot{h}_+(U_1) h_{\times} (U_1) 
			- \frac{1}{4}\int_{U_0}^{U} dU_1 \ddot{h}_{\times} (U_1)h_+(U_1) \right) \\
			&- \frac{V_0}{\sqrt{2}}\bigg[-\frac{1}{2}h_\times(U) \\
			& +\frac{1}{4}\int_{U_0}^{U}dU_1\ddot{h}_\times(U_1)h_+(U_1)(U_1-U_0) -\frac{1}{2}\int_{U_0}^{U}dU_1\ddot{h}_\times (U_1)\int_{U_0}^{U_1}dU_2h_+(U_2) \\
			& -\frac{1}{4}\int_{U_0}^{U}dU_1\ddot{h}_+(U_1)h_\times (U_1)(U_1-U_0) +\frac{1}{2}\int_{U_0}^{U}dU_1\ddot{h}_+(U_1)\int_{U_0}^{U_1}dU_2h_\times (U_2)\bigg ]\,.
		\end{aligned}
	\end{equation}
	
	\begin{figure}[h!]
		\centering
		\includegraphics[width=0.8\textwidth]{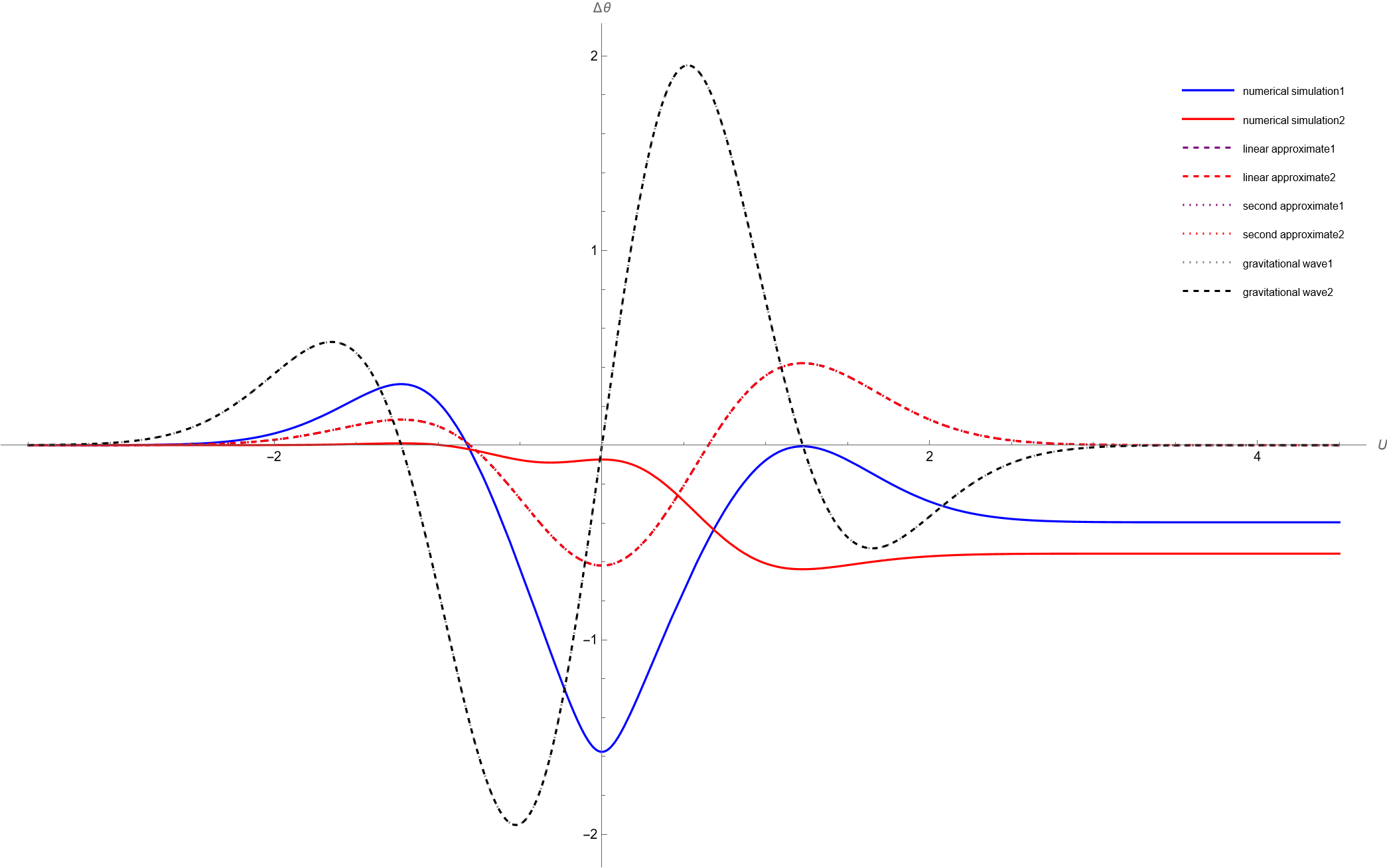}
		\caption{
			The angular deviation between two gyroscopes evolves with the initial separation velocity, accounting for X-polarization. 
			This figure shows the results of a numerical simulation based solely on a toy model of gravitational collapse, 
			neither the first-order expansion nor the second-order expansion is shown to exhibit the memory effect. The memory effect is contributed by those of higher orders. \label{fig:mathematica_plot13}}
	\end{figure}
	
	\begin{figure}[h!]
		\centering
		\includegraphics[width=0.8\textwidth]{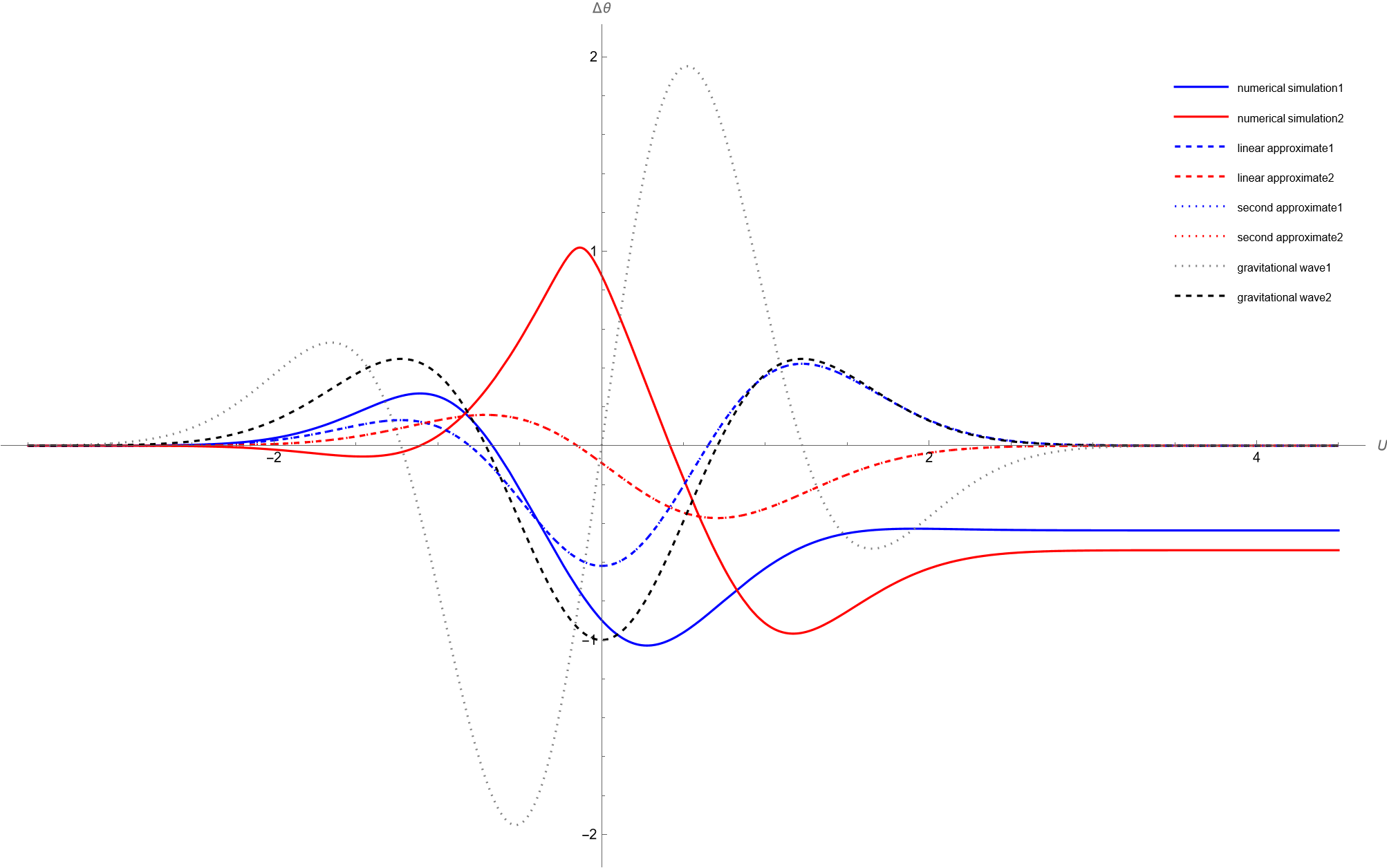}
		\caption{The evolution of the angular deviation between two gyroscopes is analyzed with respect to the initial separation velocity, incorporating X polarization. In this context,+ polarization is examined within the framework of the \eqref{eq:L3} model.
		Numerical simulations show that the contributions of both \(+\) polarization and \(\times\) polarization to the memory effect are in the higher orders. \label{fig:mathematica_plot20}}
	\end{figure}
	
	\section{The memory effect of compact binary sources}\label{sec:com}
	In a binary system, at the leading-order approximation, the memory contribution from \(\times\) polarization is zero. 
	However, when higher-order modes are included, a non-zero memory effect from \(\times\)  polarization arises in all cases except for equal-mass, non-spinning binaries.
	This higher-order effect introduces a correction to the memory signal, modifying the leading-order memory by a factor of \(10\%\)\cite{talbot2018gravitational}.
	At the 2.5PN approximation, the contribution from \(h_{\times}\) is no longer \(0\) but is instead given by \cite{nichols2017spin,arun20052}:
	\begin{equation}
		\begin{aligned}
			h_{\times}^{smm}=\frac{12M\eta^2}{5r}x^{\frac{7}{2}}sin^2\theta cos\theta+O(c^{-2})\,,
		\end{aligned}
	\end{equation}
	where M is the total mass,alongside \(\eta =\frac{m_1m_2}{M^2}\), \(x=(M\omega)^{\frac{2}{3}}\).
	Considering \(\times\) polarization and initial separation conditions that include both a displacement of \(L\) and an initial velocity of \(v_0\),the first-order approximation in standard unit is obtained as follows:
	\begin{eqnarray}
			\Delta \Theta^2  &=&\frac{v_0}{2\sqrt{2}c}h_+(U)\,, \\
			\Delta \Theta^3  &=&\frac{v_0}{2\sqrt{2}c}h_{\times}(U)\,, \\
			\Delta \Theta &=&\frac{v_0}{2\sqrt{2}c}\sqrt{h_+^2+h_{\times}^2}\,.   \label{eq:xV}
	\end{eqnarray}
	When \(\times\) polarization is considered, and the initial separation velocity approaches the speed of light, the order of magnitude for \(\Delta \Theta\) is \(10^{-16}\)rad.
	
	\section{Conclusions }\label{sec:conclusion}
	In this paper, the general expressions for the initial separation distance and separation velocity of two separated spin vectors in the plane wave spacetime are calculated by means of the formal solution of the geodesic equation. 
	Although the geodesic equation can be directly employed for calculation, we are inclined to use the formal solution for calculation due to numerous properties of sturm-Liouvile matrix equation \(\bm{\ddot{P}}=\bm{K}\bm{P}\) (see details in \cite{zhang2018sturm}).
	
	In the plane wave spacetime, the deviation angle between two spin vectors is found to involve only velocity. In the shackwave spacetime, the angular deviation resulting from the separation of two spin vectors is also calculated to correspond to the velocity memory effect(see Appendix A).
	Under the premise of linear polarization, when the sandwich wave is utilized for numerical simulation, neither the first-order expansion of \(P\) nor that of \(M\) presents the velocity memory effect. This is due to the fact that the selected sandwich wave model, 
	namely the gravitational collapse model, is symmetrical in itself and remains so after a single integration. The velocity memory effect is manifested only when the symmetry is disrupted.
	For the gravitational collapse model, when only the initial separation distance (or the initial separation velocity) is taken into account, the angle corresponding to the second-order expansion of \(\bm{P}\) (or \(\bm{M}\)) is shown to exhibit the velocity memory effect.
	The gravitational wave memory effect stems from the non-oscillatory components of the gravitational wave, and the angular deviation within the sandwich wave model is also induced by the non-oscillatory components.
	
	Calculations considering the initial separation distance and velocity between two test masses are deemed highly necessary. In a gravitational wave detector, a certain distance of separation exists between the two test masses, and initially, 
	they are not in a static state due to the influence of other forces. Based on the estimation of the plane wave approximation by \cite{divakarla2021first}, the arm length of the detector is required to be \(L\ll 10^{17}\) meters. Pulsars are preferred to be utilized and regarded as gyroscopes\cite{seraj2023gyroscopic}. 
	The order of magnitude of the contribution from the initial separation distance \(L\) of the merger of supermassive black holes to the angle is \(10^{-23}\) rads.
	It is estimated by \cite{wang2023spin} that when the initial separation velocity of two gyroscopes approaches the speed of light, the maximum angular deviation of the two gyroscopes is \(10^{-14}\) rads. 
	If the initial separation velocity of the two gyroscopes is \(V_0\ll c\), then the first-order contribution of the initial separation velocity is found to have the same order of magnitude as the leading term contribution of the aforementioned initial separation distance.

	Apart from the equal-mass, non-spinning binary systems, a non-zero \(\times\) polarization component  is exhibited by all the considered systems upon the inclusion of higher-order modes\cite{talbot2018gravitational}. Hence, the velocity memory effect with both \(+\) polarization and \(\times\) polarization taken into account is calculated. 
	The first-order term corresponds to \eqref{eq:xV}. In the case of considering the merger of supermassive black holes and with the initial separation velocity approaching the speed of light, the order of magnitude of the angle is \(10^{-16}\) rads.
	
	\appendix
	\renewcommand{\thesection}{Appendix \Alph{section}}
	\numberwithin{equation}{section}
	\section{Shackwave spacetime}
	In Brinkmann coordinates \((U,V,x^1,x^2)\), the metric of the shackwave spacetime is given as\cite{gray2021quantum}: 
	\begin{equation}
		\begin{aligned}
			ds^2=-dUdV+f(\vec{x})\delta(U-U_0)dU^2+\delta_{ij}dx^idx^j \,.
		\end{aligned}
	\end{equation}
	An orthonormal tetrad frame is constructed as follows: 
	\begin{equation}
		\begin{aligned}
			e_{\hat{0}}&=\vec{u} \,, \\
			e_{\hat{1}}&=\frac{-2\vec{l}}{\gamma}+\vec{u} \,, \\
			e_{\hat{a}}&=\partial_a+2v_a\vec{l} \,,
		\end{aligned}
	\end{equation}
	where \(\vec{l}=\partial_V\)
	The four velocity in Brinkmann coordinates can be generally written as \(\vec{u}=\gamma(1,v^1,v^a)\)
	with  \(\vec{u}\cdot \vec{u}=-1\)  and  \(\gamma=(v_1-f(\vec{x})\delta(u-u_0)+(v^a)^2)^{-\frac{1}{2}}\)
	and the result of the calculation is:\\
	\begin{equation}
		\begin{aligned}
			\omega^{(a)(b)}_{\nu}&=0 \,,\\
			\omega^{\hat{1}\hat{a}}_{\nu}&=-\gamma\partial _{\nu}v^a-\frac{\gamma}{2}\Gamma^1_{b\nu}\delta^{ab} \,, \\
			\Omega^{\hat{1}\hat{a}}&=-u^{\nu}\omega^{\hat{1}\hat{a}}_\nu=u^{\nu}\gamma\partial_{\nu}v^a+u^{\nu}\frac{\gamma}{2}\Gamma^{1}_{b\nu}\delta_{ab}\,. \\
		\end{aligned}
	\end{equation}
	When only the acceleration in the \(X^1\) direction is considered, the following holds:\\
	\begin{equation}
		\begin{aligned}
			\Omega_{\hat{1}\hat{2}}&=u^{\nu}\gamma\partial_{\nu}v^2=\gamma\frac{\mathrm{d} v^2}{\mathrm{d} \tau }  \,\\
			\theta^a& =\int_{0}^{v^a_{mem}} (1-(v)^2)^{-\frac{1}{2}}dv=\arcsin(v^a_{mem}) \,.
		\end{aligned}
	\end{equation}
	That is, the angular difference resulting from the separation of two spin vectors in the shackwave spacetime is found to correspond to the velocity memory effect.

	\section{Acknowledgments}
	  The authors acknowledge the support from NSFC grants No.11105091. 

	\bibliographystyle{unsrt}
	\bibliography{refs}

\begin{thebibliography}{10}

\bibitem{wang2023spin}
Ke~Wang and Chao-Jun Feng.
\newblock Spin vector deviation and the gravitational wave memory effect between two free-falling gyroscopes in the plane wave spacetimes.
\newblock {\em Physical Review D}, 107(8):084044, 2023.

\bibitem{zel1974radiation}
Ya~B Zel'Dovich and AG~Polnarev.
\newblock Radiation of gravitational waves by a cluster of superdense stars.
\newblock {\em Soviet Astronomy, Vol. 18, p. 17}, 18:17, 1974.

\bibitem{christodoulou1991nonlinear}
Demetrios Christodoulou.
\newblock Nonlinear nature of gravitation and gravitational-wave experiments.
\newblock {\em Physical review letters}, 67(12):1486, 1991.

\bibitem{blanchet1992hereditary}
Luc Blanchet and Thibault Damour.
\newblock Hereditary effects in gravitational radiation.
\newblock {\em Physical Review D}, 46(10):4304, 1992.

\bibitem{grishchuk1989gravitational}
LP~Grishchuk and AG~Polnarev.
\newblock Gravitational wave pulses with ‘velocity coded memory’.
\newblock {\em Sov. Phys. JETP}, 69:653, 1989.

\bibitem{zhang2018velocity}
P-M Zhang, C~Duval, GW~Gibbons, and PA~Horvathy.
\newblock Velocity memory effect for polarized gravitational waves.
\newblock {\em Journal of Cosmology and Astroparticle Physics}, 2018(05):030, 2018.

\bibitem{herrera2000influence}
L~Herrera and JL~Hernandez Pastora.
\newblock On the influence of gravitational radiation on a gyroscope.
\newblock {\em Classical and Quantum Gravity}, 17(18):3617, 2000.

\bibitem{seraj2023gyroscopic}
Ali Seraj and Blagoje Oblak.
\newblock Gyroscopic gravitational memory.
\newblock {\em Journal of High Energy Physics}, 2023(11):1--24, 2023.

\bibitem{faye2024gyroscopic}
Guillaume Faye and Ali Seraj.
\newblock Gyroscopic gravitational memory from quasi-circular binary systems.
\newblock {\em Classical and Quantum Gravity}, 2024.

\bibitem{zhang2017soft}
P-M Zhang, C~Duval, GW~Gibbons, and PA~Horvathy.
\newblock Soft gravitons and the memory effect for plane gravitational waves.
\newblock {\em Physical Review D}, 96(6):064013, 2017.

\bibitem{duval2017carroll}
Christian Duval, GW~Gibbons, PA~Horvathy, and PM~Zhang.
\newblock Carroll symmetry of plane gravitational waves.
\newblock {\em Classical and Quantum Gravity}, 34(17):175003, 2017.

\bibitem{harte2015optics}
Abraham~I Harte.
\newblock Optics in a nonlinear gravitational plane wave.
\newblock {\em Classical and Quantum Gravity}, 32(17):175017, 2015.

\bibitem{audagnotto2024dynamics}
Giulio Audagnotto and Antonino Di~Piazza.
\newblock Dynamics, quantum states and compton scattering in nonlinear gravitational waves.
\newblock {\em Journal of High Energy Physics}, 2024(6):1--24, 2024.

\bibitem{bieri2014perturbative}
Lydia Bieri and David Garfinkle.
\newblock Perturbative and gauge invariant treatment of gravitational wave memory.
\newblock {\em Physical Review D}, 89(8):084039, 2014.

\bibitem{talbot2018gravitational}
Colm Talbot, Eric Thrane, Paul~D Lasky, and Fuhui Lin.
\newblock Gravitational-wave memory: waveforms and phenomenology.
\newblock {\em Physical Review D}, 98(6):064031, 2018.

\bibitem{divakarla2021first}
Atul~K Divakarla and Bernard~F Whiting.
\newblock First-order velocity memory effect from compact binary coalescing sources.
\newblock {\em Physical Review D}, 104(6):064001, 2021.

\bibitem{flanagan2020persistent}
{\'E}anna~{\'E} Flanagan, Alexander~M Grant, Abraham~I Harte, and David~A Nichols.
\newblock Persistent gravitational wave observables: Nonlinear plane wave spacetimes.
\newblock {\em Physical Review D}, 101(10):104033, 2020.

\bibitem{zhang2018sturm}
P-M Zhang, M~Elbistan, GW~Gibbons, and PA~Horvathy.
\newblock Sturm--liouville and carroll: at the heart of the memory effect.
\newblock {\em General Relativity and Gravitation}, 50:1--9, 2018.

\bibitem{zhang2017memory}
P-M Zhang, Christian Duval, GW~Gibbons, and PA~Horvathy.
\newblock The memory effect for plane gravitational waves.
\newblock {\em Physics Letters B}, 772:743--746, 2017.

\bibitem{aggarwal2019nanograv}
K~Aggarwal, Z~Arzoumanian, PT~Baker, A~Brazier, MR~Brinson, PR~Brook, S~Burke-Spolaor, S~Chatterjee, JM~Cordes, NJ~Cornish, et~al.
\newblock The nanograv 11 yr data set: limits on gravitational waves from individual supermassive black hole binaries.
\newblock {\em The Astrophysical Journal}, 880(2):116, 2019.

\bibitem{nichols2017spin}
David~A Nichols.
\newblock Spin memory effect for compact binaries in the post-newtonian approximation.
\newblock {\em Physical Review D}, 95(8):084048, 2017.

\bibitem{arun20052}
KG~Arun, Luc Blanchet, Bala~R Iyer, and Moh’d S~S Qusailah.
\newblock The 2.5 pn gravitational wave polarizations from inspiralling compact binaries in circular orbits.
\newblock {\em Classical and Quantum Gravity}, 22(14):3115, 2005.

\bibitem{gray2021quantum}
Finnian Gray, David Kubiz{\v{n}}{\'a}k, Taillte May, Sydney Timmerman, and Erickson Tjoa.
\newblock Quantum imprints of gravitational shockwaves.
\newblock {\em Journal of High Energy Physics}, 2021(11):1--32, 2021.

\end{thebibliography}
	
\end{document}